\documentclass[a4paper,11pt]{article}
\pdfoutput=1 % if your are submitting a pdflatex (i.e. if you have
\usepackage{jcappub}
\usepackage[T1]{fontenc} % if needed
\usepackage{slashed,changes}

\newcommand{\GeV}{\mathrm{GeV}}
\newcommand{\TeV}{\mathrm{TeV}}
\newcommand{\PeV}{\mathrm{PeV}}

\usepackage{xcolor,cancel}

\title{\boldmath Linking the KM3-230213A Neutrino Event to Dark Matter Decay and Gravitational Wave signals}

\author[a,1]{Sarif Khan}
\author[a,1]{, Jongkuk Kim \note{correspondence} }
\author[b]{, and Pyungwon Ko}
 
\affiliation[a]{Department of Physics, Chung-Ang University, Seoul 06974, Korea}
\affiliation[b]{School of Physics, Korea Institute for Advanced Study,  Seoul 130-722, Korea}
 
\emailAdd{sarifkhan@cau.ac.kr}
\emailAdd{jongkukkim@cau.ac.kr}
\emailAdd{pko@kias.re.kr}

\abstract{ 
The KM3NeT collaboration recently reported the detection of an ultra-high-energy (UHE) neutrino 
event, dubbed KM3-230213A. This is the first observed neutrino event with energy of the order of 
$\mathcal{O}(100)\,\PeV$, the origin of which remains unclear.  
In this paper, we interpret this high energy neutrino event in terms of the Dirac fermion dark 
matter (DM) $\chi$ decays via the right-handed (RH) neutrino portal assuming the 
Type-I seesaw mechanism for neutrino masses and mixings. 
Furthermore, the Dirac fermion dark matter $\chi$  is assumed to be charged under $U(1)_X$ dark gauge symmetry, which is spontaneously 
broken by the vacuum expectation value (VEV) of the dark Higgs $\Phi$.
In this scenario, DM can decay into a pair of Standard Model (SM) particles, such as neutrinos, leptons, and gauge bosons via the RH neutrino portals for $v_\Phi \gg m_\chi$.  
Then we can reply on the HDMSpectra package to generate the neutrino and $\gamma$-ray spectra from heavy DM decays.
If the DM mass is around $440\, \PeV$ with a lifetime $5\times 10^{29}$ sec, 
it can account for the KM3-230213A event.
However, such heavy DM cannot be produced through the thermal freeze-out mechanism due to 
overproduction and violation of unitarity bounds.
We focus on the UV freeze-in production of DM through a dimension-5 operator, which 
helps in producing the DM dominantly in the early Universe.
Finally, the large value of the dark Higgs field VEV opens up the intriguing possibility of generating gravitational waves (GWs) spectra from cosmic strings.
We have found a reasonable set of parameter values that can address the KM3NeT signal, 
yield the correct value of the DM relic density through freeze-in mechanism, 
and allow for the possible detection of GW signal at the future detectors.
 }

\hypersetup{
colorlinks = true,
linkcolor = blue,
citecolor = magenta
}

\begin{document}
\maketitle
\flushbottom

%----------------------------------------------------------------------------------
\section{Introduction}\label{sec:intro}
%----------------------------------------------------------------------------------
The KM3NeT collaboration recently announced the detection of a muon event with an energy of $\mathcal{O}(100~\mathrm{PeV})$, originating from a muon neutrino.
This event, which now sets the record for the highest-energy neutrino ever detected, has been named KM3-230213A.
The inferred energy of the ultra-high-energy neutrino lies in the range $72~\mathrm{PeV} \leq E_\nu \leq 2.6~\mathrm{EeV}$, with a median energy of $220~\mathrm{PeV}$.
The neutrino flux required to fit the data is around \cite{KM3NeT:2025npi}
\begin{align}
	E^2_\nu \frac{d\Phi_\nu}{dE_\nu} &=	5.8_{-3.7}^{+10.1} \times 10^{-8} \GeV {\rm cm^{-2} s^{-1} sr^{-1}},
    \label{KM3flux}
\end{align}
with $E_\nu =72$ PeV -- $2.6$  EeV.

To produce such an UHE neutrino via cosmic-ray interactions (such as $pp$, $p\pi^-$, or $p\gamma$ collisions) in typical astrophysical sources would require protons accelerated to EeV energies.
However, the KM3NeT collaboration investigated several potential source candidates such as blazars and gamma-ray bursts using various electromagnetic telescope catalogs \cite{KM3NeT:2025npi}.
No conclusive evidence for a counterpart was found along the direction of the neutrino.
To explain the event, there are a lot of hypotheses proposed in Refs. \cite{Shimoda:2024qzw, Li:2025tqf, Muzio:2025gbr, KM3NeT:2025vut, Dzhatdoev:2025sdi, Barman:2025bir, Podlesnyi:2025aqb, Neronov:2025jfj, DeLaTorreLuque:2025zsv, Boccia:2025hpm, Brdar:2025azm, Borah:2025igh, Kohri:2025bsn, Narita:2025udw, Simeon:2025gxd, Jiang:2025blz, Alves:2025xul, Wang:2025lgn, Jho:2025gaf, Klipfel:2025jql, Choi:2025hqt,  Barman:2025hoz,Murase:2025uwv}.

The spatial distribution of DM and interaction properties between DM and SM particles still 
remain crucial issues in both cosmology and particle physics. 
In particular, the nature of its interaction like gravitational or otherwise is essential 
for understanding its role in the formation of large scale
structures and the evolution 
of the Universe. 
In this study, we consider a scenario in which dark matter is extremely heavy and have a lifetime much longer than the age of the Universe. 
Even though the lifetime is much longer than the age of the Universe, at present Universe, 
DM can decay into SM particles, including active neutrinos.
If DM mass is above $\mathcal{O}(100)~\TeV$, then it cannot be produced via thermal freeze-out.
We will consider an alternative DM production mechanism.
There is one unique feature of DM explanation for the KM3NeT neutrino event. 
There should be a sharp energy cut-off in the neutrino spectrum, which can be tested by future IceCube and KM3NeT data. 

The detection of the KM3NeT event and the non-detection of any events at IceCube, which has a larger effective area, is an important issue to be resolved. 
In Ref.~\cite{Li:2025tqf}, authors have shown a $3.5\sigma$ discrepancy between the KM3NeT and IceCube data assuming that the neutrino comes from a diffuse isotropic neutrino source. 
Additionally, they have further shown that the tension lies in the $3.1 - 3.6\sigma$, $2.9\sigma$, and $2\sigma$ ranges, depending on the neutrino source considered, such as cosmogenic sources, steady point and transient point sources, respectively. 
KM3NeT has also pointed out that an $2.2\sigma$ upward fluctuation is required to be consistent with the nondetection of such high-energy neutrinos at IceCube. 
In Ref.~\cite{Brdar:2025azm}, a sterile-active neutrino mixing was taken into account to resolve the tension. Authors of Ref. \cite{Brdar:2025azm} have shown that when a high-energy neutrino at the $\mathcal{O}(keV)$ scale passes through rock, it can oscillate into an active neutrino, thereby explaining the KM3NeT signal. This mechanism can be readily applied in our study by considering the two right-handed neutrinos, which contribute to neutrino mass in the $\mathcal{O}(keV-MeV)$ range and are produced from the decay of the PeV-scale DM. We have not explored this possibility in the present work.
Moreover, the simple diffuse astrophysical power-law flux of neutrinos leads to a $\mathcal{O}(3\sigma)$ discrepancy between the observed signal at KM3NeT and the non-observation at the IceCube and Auger experiments. In the present work, we have estimated the neutrino flux from DM decay, which depends on the DM mass and its lifetime. In Ref.~\cite{Kohri:2025bsn}, the authors have shown that the DM mass range $1.52\times 10^{8}$ GeV-$5.2 \times 10^{9}$ GeV and its lifetime in the range $1.42\times 10^{30}$ sec-$5.4\times 10^{29}$ sec can explain the KM3NeT signal while also predicting a flux below the IceCube limit when DM decay is neutrino-philic. Furthermore, with the aforementioned ranges, Ref.~\cite{Kohri:2025bsn} also points out that the discrepancy between the observation of the PeV neutrino signal at KM3NeT and the non-observation at IceCube and Auger is reduced to $1.2\sigma$ when the decaying DM scenario is considered. Although our study is based on a similar strategy as Ref.~\cite{Kohri:2025bsn} for explaining the KM3NeT signal from the decaying DM setup, the phenomenology is completely different.

In this work, we propose that the KM3-230213A UHE neutrino event can be due to heavy Dirac fermion dark matter $\chi$ decays into the $h\nu$ and its Goldstone--equivalent channels, 
$\chi\rightarrow Z \nu, W^\pm l^\mp$, through the right-handed (RH) neutrino portal.
DM $\chi$ is a SM-singlet, but is assumed to be charged under the local dark $U(1)_X$ gauge 
symmetry that is spontaneously broken by the nonzero VEV of dark Higgs field $\Phi$.
We consider a heavy Dirac fermion $\chi$ as a DM candidate, whose very late decay into SM 
particles is responsible for the KM3-230213A event.
Moreover, the added RH neutrinos can generate the neutrino mass through the type-I seesaw mechanism, which explains the origin of light neutrino masses. 
It is worth mentioning that among the three RH neutrinos, two account for the light neutrino masses\footnote{The masses of two of the three right-handed neutrinos can be small ($\mathcal{O}$(keV$-$MeV)), allowing them to reproduce the correct oscillation parameters (Refs. \cite{Ko:2014bka, Biswas:2018sib} discusses the successful reproduction of neutrino data with two right-handed neutrinos.). With a tiny value of the $\kappa$ parameter (defined in Section \ref{sec:model}), DM can produce highly energetic sterile neutrinos, which may oscillate into active neutrinos while propagating through rock, potentially explaining the KM3NeT signal, similar to Ref. \cite{Brdar:2025azm}.}, while the remaining one assists in the long-lived DM decay and its production through an effective operator 
generated by super-heavy RH neutrino.
Since a very high value of the VEV $v_{\phi}$ is required in enhancing the SM two-body decays 
of DM over the three-body decays, which demands $v_{\phi} \gg M_{\chi}$, where $M_{\chi}$ is the DM mass, typically in the $\mathcal{O}(100)$ PeV scale for explaining the KM3NeT signal.
Such a heavy dark Higgs VEV leads to detection prospects of the present scenario at future GW detectors.
Moreover, we have also produced such heavy DM candidates by the UV freeze-in mechanism through an effective higher-dimensional operator. We have found that such a high-mass regime can be easily achieved for such high values of the VEV without fine-tuning the model
parameters. In other words, we can explain the DM relic density and lifetime using the model 
parameters described in Sec. \ref{sec:model}, with $T_{R} \sim 10^{10}$ GeV, 
$\kappa \sim 10^{-4}$, $M_{N} \sim 10^{12}$ GeV, and $y \sim 10^{-22}$. 
The parameter values are natural for the freeze-in kind of DM production, and a tiny value of $y$ is needed to ensure a DM lifetime of $\mathcal{O}(10^{29})$ sec,
which is typical for all DM scenarios requiring a decaying DM particle.

In the present work, we have produced the neutrino flux
from the two body decay of the DM which is larger than the three body decay
for dark higgs VEV larger than the DM mass \cite{Ko:2015nma}. 
We have used the HDMSpectra package \cite{Bauer:2020jay} for
producing the neutrino and photon fluxes. With the 440 PeV DM mass and $\sim 5\times 10^{29}$
sec lifetime, we can successfully explain the KM3NeT data, 
whereas the photon flux obtained
from the same decay is lower than the LHAASO$-$KM2A and EAS$-$MSU data. 
DM dominantly decays to two-body SM final states ($h\nu, W^{\pm}l^{\mp}, Z\nu $), and we have produced the neutrino and photon flux from there using HDMSpectra \cite{Bauer:2020jay}. 
Since neutrinos are directly involved in the final states and photons come from the cascade decays, 
we have the dominant production of the neutrino flux compared to the photon flux. 
%In a scenario where DM only decays as $\chi \rightarrow Z \nu, W^{\pm}l^{\mp}$, we expect compared to neutrino flux, photon flux are dominant because of the larger hadronic branchings of $W^{\pm}, Z$. %due to the leptonic branchings. 
%In our case two-body decays directly produce neutrinos. 
Therefore, the neutrino flux is dominating compared to the photon flux spectrum.

In the present work, the $U(1)_X$ dark gauge symmetry breaks spontaneously 
at a very high scale nonzero VEV of dark Higgs field $\Phi$.  
Then the cosmic string networks can form during the 
dark gauge symmetry breaking.  The cosmic string production in the early Universe as a remnant of the 
symmetry-breaking phase transition has been predicted in a wide variety of 
Grand Unified Theories (GUT) as well \cite{Antusch:2023zjk, Antusch:2024nqg, Maji:2024tzg}. 
The $U(1)_X$ gauge symmetry spontaneously breaks down, 
resulting in a non-trivial value for the first homotopy group; hence, cosmic strings 
form during the symmetry breaking. Once the cosmic string networks are formed, they 
achieve a scaling regime and produce long loops that emit gravitational waves (GW) and 
vanish over time.

Cosmic strings produced from the spontaneous symmetry breaking of the $U(1)_X$ gauge symmetry 
have a string tension related to the VEV of the BSM scalar as $\mu \sim v^2_{\phi}$ \cite{Kibble:1976sj}. In the scaling regime, the fraction of cosmic string networks energy budget compared to the total energy budget is proportional to string tension $\mu$.
Therefore, a very high value of the $U(1)_X$-breaking scalar VEV can be probed at different proposed detectors,  namely  Square Kilometre Array (SKA) \cite{Janssen:2014dka}, 
Laser Interferometer Space Antenna (LISA) \cite{LISA:2017pwj},  Big Bang Observer 
(BBO)/DECi hertz Interferometer Gravitational wave Observatory (DECIGO) \cite{Yagi:2011wg},  
Einstein Telescope (ET) \cite{Punturo:2010zz, Hild:2010id},
Cosmic Explorer (CE) \cite{LIGOScientific:2016wof}, and 
LIGO/Virgo \cite{LIGOScientific:2017adf, LIGOScientific:2014qfs}. 
As we will see, $G\mu > 2\times 10^{-11}$ ($v _{\phi} > 10^{14}$ GeV) 
\footnote{$G$ is the gravitational constant.}
is already in tension with the  European Pulsar Timing Array (EPTA) data \cite{EPTA:2011kjn}.
Moreover, we will see that $G\mu < 10^{-19}$ is beyond the reach of the proposed GW detectors.

This paper is organized as follows. 
In Sec.~\ref{sec:model} we introduce our DM model with dark $U(1)_X$ gauge symmetry, heavy right-handed neutrino portal and Higgs portal interactions. 
In Sec.~\ref{sec:spectrum} we outline the general formalism for calculating neutrino flux from galactic and extragalactic DM decay. 
We present both total and differential decay widths for the relevant three-body decay in our model and compare the numerical results with IceCube data. 
In Sec.~\ref{sec:relic} we discuss a possible mechanism to generate the correct relic density for DM within our scenario of decaying heavy Dirac fermion DM and detection constraints. 
In Sec.~\ref{sec:GW}, we discuss gravitational signature induced by cosmic strings.
Finally, we conlcude in Sec.~\ref{sec:conclusion}. 

%----------------------------------------------------------------------------------
\section{Dark $U(1)_X$ model}\label{sec:model}
%----------------------------------------------------------------------------------
In this section, we recapitulate the decaying heavy Dirac fermion DM model that was proposed in Ref.~\cite{Ko:2015nma} in order to describe high energy neutrino events reported by IceCube.
Let us start with dark $U(1)_X$ gauge symmetry including a Dirac fermion DM $\chi$ and a dark Higgs field $\Phi$ \cite{Ko:2015nma}. 
Their charge assignments under the dark $U(1)_X$ symmetry are as follows: $(Q_\chi, Q_\Phi) = (1,1) .$
We consider the renormalizable and gauge-invariant Lagrangian that includes singlet right-handed 
(RH) neutrinos $N$'s which are gauge singlets: 
\begin{eqnarray}
\mathcal{L} &=& \mathcal{L}_{\mathrm{SM}} + \frac{1}{2}\bar{N}i\slashed{\partial}N - 
\left(\frac{1}{2}m_{N} \bar{N}^c N + y \bar{L}\widetilde{H} N  + \textrm{h.c.}\right) 
-\frac{1}{4}X_{\mu\nu}X^{\mu\nu} -\frac{1}{2}\sin{\epsilon}X_{\mu\nu}F^{\mu\nu}_{Y}  \nonumber\\ 	
&& + D_\mu \Phi^{\dagger}D^\mu \Phi -V(\Phi,H)
+ \bar{\chi}\left(i\slashed{D} - m_\chi\right)\chi  - \left(\kappa \bar{\chi}\Phi N  +\textrm{h.c.}\right), 
\end{eqnarray}
where $L = (\nu , l)^T$ denotes the left-handed (LH) SM lepton doublet, and $H$ is the SM Higgs doublet.  The field strength tensor of the $U(1)_X$ gauge field $X_\mu$ is given by $X_{\mu\nu} = \partial_\mu X_\nu - \partial_\nu X_\mu$, while $F^{\mu\nu}_Y$ corresponds to that of the SM 
gauge field for $U(1)_Y$ hypercharge.
The parameter $\epsilon$ represents the kinetic mixing between the SM hypercharge and the new $U(1)_X$ gauge boson.
The covariant derivative is defined as $D_\mu = \partial_\mu - i g_X Q_X X_\mu$.
Here we introduce two new types of Yukawa couplings, $y$ and $\kappa$, which are assumed to be real for simplicity. 
Note that flavor indices on $L, N$ are suppressed for simplicity, and $y$ is the usual Dirac Yukawa 
couplings in the Type-I seesaw models. In the present work, among the three right-handed neutrinos, two contribute to the neutrino mass with tiny $\kappa$ values, whereas the other right-handed neutrino has $\kappa \sim \mathcal{O}(10^{-3})$ and takes part in DM production and the KM3NeT signal, with a negligible contribution to the neutrino mass.

The scalar potential $V(\Phi, H)$, including the dark Higgs field $\Phi$, is given by
\begin{equation}
	V(\Phi, H)=\lambda_H \left(H^\dagger H -\frac{v^2_H}{2}\right)^2+\lambda_{\phi H}\left(H^\dagger H -\frac{v^2_H}{2}\right) \left(\Phi^\dagger \Phi -\frac{v^2_\phi}{2}\right) + \lambda_\phi \left(\Phi^\dagger \Phi -\frac{v^2_\phi}{2}\right)^2.
\end{equation}
Both electroweak (EW) and dark gauge symmetries are spontaneously broken by the nonzero 
vacuum expectation values of $H$ and $\Phi$, respectively.
In the unitarity gauge, the scalar fields read
\begin{equation}
	H (x) = \frac{1}{\sqrt{2}}
	\left(
	\begin{array}{c}
		0\\ v_{H}+h (x)
	\end{array}
	\right)
	~~ {\rm and}~~
	\Phi (x) = \dfrac{v_\phi+\phi (x)}{\sqrt{2}} .
\end{equation}
Two electrically neutral scalars $h$ and $\phi$ can mix with each other due 
to the Higgs-portal coupling, $\lambda_{\phi H}$, forming two mass eigenstates, $H_1 \simeq h$ and 
$H_2 \sim \phi$.  Thanks to this mixing, dark Higgs $\phi (\simeq H_2)$ can decay into SM particles. 

Likewise, three neutral gauge bosons, photon $A_\mu$, $Z_\mu$ and $X_\mu$ can mix 
with each other due to the kinetic mixing $\epsilon$.
Here we take the kinetic mixing to be small, $\epsilon \ll 1$. 
New physical gauge boson $Z'_\mu$ in the mass eigenstate is mostly 
dark photon $X_\mu$ which can decay SM fermion pairs.

When the RH neutrino $N_R$ is significantly heavier than the dark matter $\chi$ 
\footnote{In this work, we assume the RH neutrinos are much heavier than the DM $\chi$: 
see Eq.~\eqref{range-parameter} for the ranges of various parameters. We assume very tiny $\kappa$ values for two of the three right-handed neutrinos, which take part in the neutrino mass and do not contribute to the KM3NeT signal. This ensures that the DM lifetime remains unchanged and the correct neutrino oscillation data are reproduced.}, 
it can be integrated out, yielding the dim--5
effective operator \cite{Ko:2015nma}:
\begin{equation}
	\frac{y\kappa}{m_N}\bar{\chi}\Phi H^\dagger L + h.c. \label{effective:Op}
\end{equation} 
This term enables the decay of $\chi$ by allowing it to couple to light particles.
However, dark matter can remain long-lived due to the superheavy RH 
neutrino masses and the appropriate choice of Yukawa coupling constants.
After the spontaneous breaking of gauge symmetries, the operator in Eq.~\eqref{effective:Op} leads to the emergence of various higher-dimensional effective operators  \cite{Ko:2015nma}:
\begin{align}\label{eq:operator}
	\dfrac{y\kappa}{2}\frac{v_\phi v_H}{m_N}\bar{\chi}\nu,\;\dfrac{y\kappa}{2}\frac{v_\phi}{m_N}\bar{\chi}h\nu ,\;
	\dfrac{y\kappa}{2}\frac{v_H}{m_N}\bar{\chi}\phi \nu,\;&\dfrac{y\kappa}{2}\frac{1}{ m_N}\bar{\chi}\phi h\nu.
\end{align}
If kinematically allowed, all of the above operators induce $\chi$ decays into various channels with fixed relative branching ratios.
Under the assumption that $\chi$ is much heavier than $\phi$, $Z'$, $h$, $Z$, and $W$, the mass operator $\bar{\chi}\nu$ in Eq.~\eqref{eq:operator} gives rise to a suppressed mixing between DM and the active neutrino.
The mixing angle $\theta$ between DM and active neutrino is approximately given by
\begin{equation}
	\theta \simeq \frac{y\kappa}{2}\frac{v_\phi v_H}{m_N m_\chi} \ . 
    \label{mixing-angle}
\end{equation}
Consequently, the gauge interactions of $\chi$ and $\nu$ induce decay modes for $\chi$ as follows:
\begin{equation}
	\chi\rightarrow Z'\nu, Z\nu, W^{\mp}l^{\pm}, \label{SM:decays}
\end{equation}
with their branching ratios being proportional to  $\sim v^2_H:v^2_\phi:2v^2_\phi$. 
The dim-4 operators $\bar{\chi}h\nu$ and $\bar{\chi}\phi \nu$ in Eq.~\eqref{eq:operator} induce the following decays of $\chi$:
\begin{equation}
	\chi\rightarrow h\nu,\phi \nu, \label{DS:decays}
\end{equation}
with their branching ratios being proportional to $\sim v^2_\phi : v^2_H$. 
Therefore,  we can evaluate all the decay branching ratios in this model.  
The branching ratios for $\chi \rightarrow h \nu , Z \nu , W^\pm l^\mp$ to be $1:1:2$, and 
$\chi \rightarrow \phi \nu, Z^{'} \nu$ to be $1:1$,  as to be understood with the Goldstone boson equivalence theorem. The decay expressions of the DM decay are shown in Eq.~\eqref{dm-decay}.

Additionally, this model permits the three-body decay $\chi \to \phi h \nu$, which can dominate depending on the mass hierarchy between $\chi$ and $v_\phi$.   We can compare two body decay channel to 
$\chi\rightarrow \phi h \nu$ three body decay channel \cite{Ko:2015nma}:
\begin{equation}
	\frac{\Gamma_2\left(\chi\rightarrow h \nu ,\phi \nu \right)}{\Gamma_3\left(\chi\rightarrow \phi h \nu \right)} \simeq 16\pi^2 \frac{v^2_\phi+v^2_H}{m^2_\chi}.
	\label{rat-2body-3body}
\end{equation}
There are another three-body decay channels: 
\begin{equation*}
	\chi \rightarrow \phi/Z' + h +\nu,\; \phi/Z' + Z + \nu,\; \phi/Z' + W^{\pm} +l^{\mp},
\end{equation*}
with branching ratios $1:1:2$ because of the Goldstone boson equivalence theorem. 

In determining the neutrino and photon fluxes, we have used the HDMspectra package \cite{Bauer:2020jay}, which provides accurate SM particle ($\bar{p} , e^\pm , \gamma , \nu$ etc.) 
spectra when dealing with the heavy particles 
decaying or pair-annihilating promptly 
into a pair of SM particles.  
In particular, we have focused on two-body decays of DM into SM particles. 
When using the HDMspectra package,  
it is more challenging to determine the spectra 
when DM decays into beyond SM particles (such as dark photon or dark Higgs) or DM 
undergoes three-body decays, 
as this requires knowledge of the branching ratios and the energies of the final-state particles, 
which can vary in the case of three-body decays. 
A dominant two-body decay into 
SM particles can be ensured by choosing $v_{\phi}\gg m_{\chi}$ 
[see Eqs.~\eqref{SM:decays} - \eqref{rat-2body-3body}]. 
Most interestingly, this choice of the large $v_\phi$ opens up a new  possibility to  
detect GWs from cosmic string networks 
with the future GW detectors in the present work (see Sec. 4).
Therefore, the most important DM decay channels 
we shall consider  are 
$\chi \to \nu h,\nu Z, \ell^\pm W^\mp $ where all the final states involve the SM particles, 
and not dark sector particles. 
The flux originating from DM decay will be detected in all active neutrino telescopes on Earth. In the present work, we reduce the discrepancy between the non-detection of high-energy neutrinos at IceCube or Pierre Auger and their detection at KM3NeT by suitably choosing the DM mass and decay lifetime.

%----------------------------------------------------------------------------------
\section{Neutrino Flux from DM decay}\label{sec:spectrum}
%----------------------------------------------------------------------------------
The neutrino flux from dark matter decay is split into two parts: galactic and extragalactic contributions.
Galactic neutrino flux at kinetic energy $E$ from DM decay in our Milky Way dark halo is given by
\begin{equation}\label{eq:flux_g}
	\left.\frac{d\Phi_\nu^G}{dE_\nu d\Omega }\right|_{E_\nu=E}= \frac{D_g}{4\pi M_\chi}\sum_{i}  \Gamma_i \left.\frac{dN^i_\nu}{d E_\nu} \right|_{E_\nu=E},
\end{equation}
where $\Gamma_i$ is partial width for decay channel $i$, $d N^i_\nu/dE_\nu$ is the neutrino spectrum at production obtained by HDMSpectra \cite{Bauer:2020jay}.
$D_g$ is the D-factor from our galaxy which is defined by
\begin{align}
	D_g &= \frac{1}{\Delta\Omega} \int_{\Delta\Omega} d\Omega\int_{0}^{r_{\rm max}} dr' \rho(r'), \label{gal:DM}
\end{align}
where $\Delta\Omega$ is the angular region of detection, $r'=\sqrt{r_\odot^2 + r^2 -2r_\odot  r \cos \psi}$, $r$ is the distance to earth from the DM decay point,  $r_\odot\simeq 8.5$ kpc is the distance between galactic center and the Solar system. 
For the galactic DM density distribution, we use the following standard NFW profile~\cite{Navarro:1995iw},
\begin{equation}\label{eq:NFW}
	\rho_\chi\left(r \right)=\frac{\rho_0}{r/r_c \left( 1+r/r_c \right)^2},
\end{equation}
with parameters $r_c\simeq 20$ kpc and $\rho_0$ is determined by  $\rho_\chi (r_\odot )= 0.3~\textrm{GeV}/\textrm{cm}^3$. 
For KM3-230213A event, the neutrino flux is not sensitive to the choice of
DM density profile in our galaxy because a neutrino does not travel from the galactic center. 
When DM comes from the galactic center, the value of the integration in Eq.~\eqref{gal:DM} is different depending on which DM profile we used.

We can also get the extragalactic or cosmic contribution, by taking cosmic expansion into account, namely the redshift effect~\cite{Cirelli:2010xx}:
\begin{align}
	\frac{d\Phi_\nu^{EG}}{dE_\nu d\Omega} &= \int_{z_0}^{\infty} dz \frac{1}{ \mathcal{H}(z) (1+z) } \left( \frac{1+z_0}{1+z}  \right)^3 \frac{\bar{\rho}_\chi (z) }{m_\chi} \sum \Gamma_i  \left.\frac{dN^i_\nu}{d E'_\nu} \right|_{E'_\nu=(1+z)E} e^{-s_\nu(E_\nu,z) } ,
\end{align}
where the average cosmological DM density is $\bar{\rho}_\chi(z) = \rho^0_\chi (1+z)^3$, the factor $\left( \frac{1+z_0}{1+z}  \right)^3$ incorporates the effect of cosmological dimming of surface brightness due to the expansion-induced dilution of the source, and the factor $\left( \mathcal{H}(z) (1+z)\right)^{-1}$ converts the redshift interval into a corresponding proper distance interval. 
Hubble parameter is given by,
\begin{align}
	\mathcal{H}(z) &= \mathcal{H}_0 \sqrt{ \Omega_\Lambda + \Omega_\mathrm{m}(1+z)^3 + \Omega_{r}(1+z)^4},
\end{align}
where $\mathcal{H}_0$ is the present-day value of the Hubble parameter.
Since we observe the neutrino at $z_0=0$, the differential flux of extragalactic neutrino reads
\begin{align}\label{eq:flux_eg}
	\frac{d\Phi_\nu^{EG}}{dE_\nu d\Omega} &=D_{\rm eg} \int_0^\infty \frac{dz}{1+z}\frac{1}{\sqrt{\Omega_\Lambda + \Omega_\mathrm{m}(1+z)^3 + \Omega_{r}(1+z)^4} } \left.\frac{dN^i_\nu}{d E'_\nu} \right|_{E'_\nu=(1+z)E} e^{-s_\nu(E_\nu,z) },
\end{align}
where $E'$ is red-shifted to $E$ as $E'=(1+z)E$. $D_{\rm eg}$ is the D-factor from extragalaxy which is defiend as
\begin{align}
	D_{\rm eg} &= \frac{ \Omega_{\chi}  \rho_c  }{4\pi M_\chi \tau_\chi \mathcal{H}_0 } = 1.6 \times 10^{-16} \left( \frac{ 100 \PeV }{ M_\chi } \right) \left(  \frac{10^{29}{\rm s} }{\tau_\chi } \right)   {\rm cm^{-2}s^{-1} sr^{-1} }, 
\end{align}
where the critical energy density $\rho_c=\frac{3 \mathcal{H}^2_0}{8 \pi G}=5.5\times 10^{-6} \GeV/\mathrm{cm}^{3}$ and $\Omega_{\chi}\simeq 0.27$ is DM $\chi$'s fraction, $\Omega_\Lambda$, $\Omega_\mathrm{m}$ and $\Omega_{r}$ are energy fractions of dark energy, total matter, and radiations, respectively. 
We have adopted the Planck results~\cite{Planck:2018vyg} for the numerical evaluation. 
Lastly, $s_\nu(E_\nu,z)$ is the neutrino opacity of the Universe, assuming that the neutrinos are massless.
The neutrino opacity $s_\nu (E_\nu, z)$ is given by \cite{WMAP:2010qai}
\begin{eqnarray}
	s_\nu (E_\nu, z) = \left\{\begin{matrix}
		7.4\times 10^{-17} (1+z)^{7/2} \left(E_\nu/ \TeV \right), ~~~\rm{for}~~ 1 \ll {\it z} < {\it z}_{\rm eq}   \\
		1.7\times 10^{-14} (1+z)^3 \left(E_\nu/ \TeV \right), ~~~~~\rm{for}~~~{\it z} \gg {\it z}_{\rm eq} ~~~~~~
	\end{matrix}\right.
\end{eqnarray}
where $z_{\rm eq} \sim 3200$ \cite{Planck:2018vyg} is the redshift value 
corresponding to the matter-radiation equality epoch during the Universe evolution\footnote{In Ref. \cite{Ema:2013nda}, the authors pointed out that damping due to the neutrino opacity parameter dominates for $z > \mathcal{O}(10^{3})$, depending on the neutrino energy. In this work, we have taken this limit as $z_{eq} \sim 3200$ using the analytical expression, and we do not expect any change even if the full solution is considered.}.
As discussed, in the present work we have considered the NFW profile for the DM distribution and accounted for the direction of the signal. We focus on the direction from which the signal was observed and include both Galactic and extra-Galactic contributions. At the signal energy, the Galactic contribution exhibits a peak, while the extra-Galactic contribution shows a subdominant peak due to redshift effects, contributing less at the signal energy. In our scenario, heavy DM is assumed as the origin of the signal, which can reduce the discrepancy between the non-detection of such a heavy neutrino signal at IceCube or Pierre Auger and its detection at KM3NeT for suitable values of DM mass and lifetime \cite{Kohri:2025bsn}.
Moreover, the signal could, in principle, arise solely from the extra-Galactic contribution if the DM lifetime is shorter than the age of the Universe and only a tiny fraction of DM is involved. This predicts no signal from the Galactic component but allows the extra-Galactic part to contribute after redshifting. Such shorter lifetimes are more consistent with Ref. \cite{KM3NeT:2025aps}, where the extra-Galactic contribution is favored due to the absence of a known Galactic astrophysical source in that direction and because the signal direction is far from the Galactic center. In the present work, however, we consider both Galactic and extra-Galactic contributions by assuming a DM lifetime greater than the age of the Universe.

Now we discuss neutrino and gamma-ray fluxes induced by decaying heavy DM.
\begin{figure}[t]
\centering
\includegraphics[height=0.42\textwidth,width=0.48\textwidth]{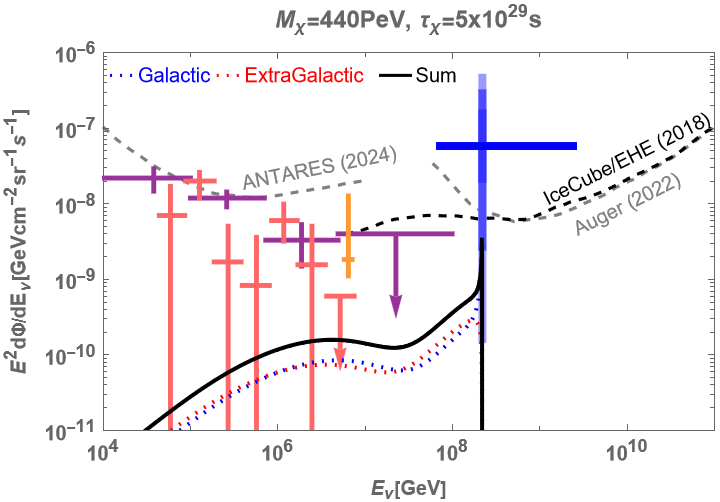}
\includegraphics[height=0.42\textwidth,width=0.48\textwidth]{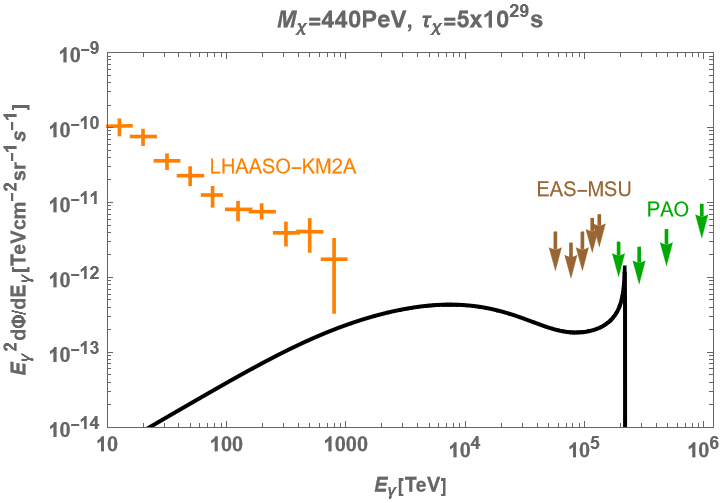} 
\caption{ Neutrino (left-panel) and  Gamma-ray (right-panel) spectra from DM $\chi$ decay with $M_\chi=440\PeV$ and lifetime $\tau_\chi =1/\Gamma = 5\times 10^{29}$s.  
In the left panel, bounds come from IceCube \cite{IceCube:2018fhm, IceCube:2020wum}. Blue cross corresponds to KM3NeT  with $3\sigma$ C.L \cite{KM3NeT:2025npi}.  
It presents the galactic (blue dotted curve) and extragalactic (red dotted curve) neutrino flux.
In the right panel, orange crosses correspond to gamma-ray constraints from LHAASO-KM2A \cite{LHAASO:2023gne} whereas EAS-MSU \cite{Fomin:2017ypo} and PAO  \cite{Castellina:2019huz} limits are shown in brown and green arrows, respectively. 
\label{fig:spectrum}
}
\end{figure}
In the left panel (LP) of Fig.~\ref{fig:spectrum}, we show 
the neutrino flux originated from both our galaxy,$(\ell,b)=(210.06^\circ, -11.13^\circ)$, and 
extragalaxy.  The neutrino flavor ratio arriving at the Earth is assumed to be $1 : 1 : 1$.
Here we consider the arrival direction (RA$:~94.3^\circ$, Dec$:~-7.8^\circ$) of KM3-230213A with an angular uncertainty of $\pm 1.5^\circ$ with $1\sigma$ C.L \cite{KM3NeT:2025npi}.
In the left panel, we describe the neutrino spectra from the heavy DM decay. 
Red crosses denote the IceCube data \cite{IceCube:2018fhm, IceCube:2020wum}.
Left dashed line represents the bound coming from the ANTARES data \cite{ANTARES:2024ihw}.
Right dashed lines above $\sim 10\PeV$ are upper limits of neutrino flux
coming from no-observations of neutrinos beyond that in IceCube and Auger 
\cite{IceCube:2018fhm, IceCube:2020wum}.
The orange cross is the most energetic neutrino event detected by IceCube neutrino telescope \cite{IceCube:2021rpz}.
We also show the preferred flux of the KM3-230213A event by Blue cross with $3\sigma$ C.L.
Blue and red dotted lines show galactic and extragalactic contributions from heavy decaying DM, respectively.  
We can explain the neutrino event detected by the KM3-NeT neutrino telescope through the decaying heavy DM. 
As shown in Eq.~\eqref{KM3flux}, KM3NeT has considered neutrino flux which varies as $E^{-2}$ with energy and determined the excess over the flux considered. In our case, the flux produced from the DM decay which is not the same flux as $E^{-2}$, but we have a dominant contribution only where the excess has been obtained, and in the rest of the region DM has a subdominant contribution to the neutrino flux and comes mainly from the flux $E^{-2}$ considered in the KM3NeT study. Moreover, in Ref. \cite{Li:2025tqf}, authors have pointed out that for neutrino energy at KM3NeT, which triggered 3672 PMTs, the peak can occur at $120$ PeV, $190$ PeV, and $335$ PeV depending on the neutrino flux considered as $E^{-2.52}$, $E^{-2}$, and cosmogenic, respectively. In the present work, by suitably choosing the DM mass and lifetime, we can explain different peaks obtained based on the flux assumptions. In general, we have dominant two-body decay, and we expect the neutrino flux has a peak around $E_{\nu} \simeq M_{\chi}/2$.

On the other hand, heavy DM $\chi$ decay can generate secondary gamma-ray flux. 
In the right panel (RP) of Fig. ~\ref{fig:spectrum}, we show the gamma-ray spectra induced 
by the heavy DM decay, including photons from the cascade decays. 
We compute the gamma-ray spectra from the inner Galactic plane, 
$15^\circ < \ell < 125^\circ, ~-5^\circ < b< 5^\circ$.
The orange crosses correspond to LHAASO-KM2A data \cite{LHAASO:2023gne}. 
Upper limites from EAS-MSU  \cite{Fomin:2017ypo} and PAO  
\cite{Castellina:2019huz} Collaborations are shown in brown and green colors, respectively.
We note that the photon flux predicted in our model is below the 
upper bounds from LHAASO-KM2A and EAS-MSU data.
We see that our photon flux is weaker than the neutrino flux. This is because we have direct neutrinos in the final states of DM decay $\chi \rightarrow h \nu$, whereas the photons come from the cascade decays of $h$, $W^{\pm}$, $l^{\pm}$. Additionally, the DM decay modes where we have $W^{\pm}$, $l^{\mp}$, and $Z$ have subdominant contribution compared to $\nu h$ due to suppression coming from the cascade decays. 
Finally, our model predicts the photon flux below the data obtained by LHAASO-KM2A and the bound put by EAS-MSU, and PAO. Therefore, in the future, our setup can be tested if any photon flux is observed at such energy scales. Moreover, DM is produced in a non-thermal way and its mass is at the PeV scale, so it could be difficult to observe in other conventional search experiments like colliders or DM direct detection. Such heavy DM is very difficult to probe directly. In Ref.\cite{Ganguly:2025ffb}, the authors have pointed out that such heavy DM can be probed through red giants after continuous capture, as gravitational collapse results in helium ignition earlier than the standard stellar evolution prediction. Moreover, in Ref.\cite{Maitra:2025opp}, the authors have pointed out that such heavy DM, when decaying to neutrinos, can interact with the cosmic neutrino background, resulting in lower-energy neutrino spectra that can be detected at IceCube-Gen2 Radio. 

%----------------------------------------------------------------------------------
\section{Gravitational waves from cosmic string}\label{sec:GW}
%----------------------------------------------------------------------------------
In this section, we discuss GW production from cosmic strings, which are formed due to the $U(1)$ gauge symmetry breaking in the early Universe. After their formation, the cosmic string network enters the scaling regime, where the strings achieve a balance between the slow redshift of the horizon-length strings due to the $a(t)^{-2}$ factor in density dilution during cosmic expansion and the transfer of energy from long strings to the production of closed string loops \cite{Vilenkin:2000jqa}. 
The velocity-dependent one-scale (VOS) model provides a good analytical description of long strings by taking into account their characteristic length and mean string velocity. A detailed description of the VOS model is given in Refs. \cite{Martins:1995tg, Martins:1996jp, Martins:2000cs}.

During the evolution of cosmic strings, large loops are produced that predominantly emit GWs, while highly boosted small loops primarily lose energy through 
simple redshifting \cite{Vanchurin:2005pa, Martins:2005es}. 
The long loops lose energy in the form of GWs at a constant rate governed by the following equation, 
\begin{eqnarray} 
	\frac{d E}{dt} = -\Gamma G \mu^2 
\end{eqnarray} 
where $\Gamma = 50$ is obtained from simulations 
\cite{Vilenkin:1981bx, Turok:1984cn, Blanco-Pillado:2017oxo}, 
$G  \equiv \frac{1}{M^2_{pl}} = 6.71\times 10^{-39} {\rm GeV}^{-2}$ is the gravitational constant, and $\mu \sim v_\phi^2$ is the string tension. Therefore, loops with an initial size $l_i = \alpha t_i$, where $t_i$ is the loop formation time and $\alpha = 0.1$ is 
an approximate loop size parameter \cite{Blanco-Pillado:2017oxo, Blanco-Pillado:2013qja}, 
shrink with time, and their length $l(t)$ at time $t$ can be expressed as, \begin{eqnarray} 
	l(t) = \alpha t_i - \Gamma G \mu (t - t_i).
\end{eqnarray}

The total energy loss from an individual loop also depends on the mode frequency $f_k = \frac{2k}{l}$ ($k = 1, 2, 3 \ldots$ is the mode number) during its normal mode oscillation. 
As found in Refs. \cite{Blanco-Pillado:2017oxo, Blanco-Pillado:2013qja}, 
the emission rate per mode scales with $k$ as $k^{-4/3}$ and takes the form, \begin{eqnarray} 
	\Gamma^{(k)} = \frac{k^{-4/3} \Gamma}{\sum_{m=1}^{\infty} m^{-4/3}} \end{eqnarray} 
where $\sum_{m=1}^{\infty} m^{-4/3} \simeq 3.60$. The frequency 
$\tilde{f}$ emitted at time $\tilde{t}$ can be related to the frequency 
$f$ at time $t$ by accounting the redshift effect as follows, \begin{eqnarray} 
	f = \left[ \frac{a(\tilde{t})}{a(t)} \right] \tilde{f}.
\end{eqnarray}

We can express the relic GW background from the cosmic string network in terms of GW energy density $\rho_{GW}$ and frequency $f$ as follows, \begin{eqnarray} 
	\Omega_{GW} (f) = \frac{1}{\rho_c} \frac{d \rho_{GW}}{d \ln f}. \end{eqnarray} 
Summing over all $k$-modes during string oscillations, the relic 
GW density for frequency $f$ can be expressed as, 
\begin{eqnarray} 
	\Omega_{GW}(f) = \sum_k \Omega^{(k)}_{GW} (f) 
\end{eqnarray} 
where the contribution to the GW from an individual frequency mode 
is given by \cite{Cui:2018rwi}, 
\begin{eqnarray} 
	\Omega_{GW}^{(k)} = \frac{1}{\rho_c} \frac{2k}{f} \frac{F_{\alpha} \Gamma^{(k)} G \mu^2}{\alpha (\alpha + \Gamma G \mu)} \int_{t_F}^{t_0} d \tilde{t} \frac{C_{eff}(t^{(k)}_i)}{(t^{(k)}_i)^4} \left[ \frac{a(\tilde{t})}{a(t_0)} \right]^5 \left[ \frac{a(t^{(k)}_i)}{a(\tilde{t})} \right]^3 \theta(t^{(k)}_i - t_F). 
	\label{GW-spectra}
\end{eqnarray}

In the above expression, $F_{\alpha} = 0.1$ denotes the fraction of energy released by long strings into loops. The parameter $t_F$ is the time when the string network reaches the scaling regime, which occurs shortly after cosmic string formation during symmetry breaking. The quantity $t_0$ denotes the present time, and $\tilde{t}$ is the time of GW emission, 
over which the integration runs from $t_F$ to $t_0$. 
The function $C_{eff}(t^{(k)}{i})$ takes the value 5.5 during the radiation-dominated era and 0.41 during the matter-dominated era in the early Universe, as obtained from 
simulation 
\cite{Blanco-Pillado:2011egf, Blanco-Pillado:2017oxo, Blanco-Pillado:2013qja, Cui:2017ufi}. 
The loop formation time contributing to mode $k$ is given by 
\begin{eqnarray} 
	t^{(k)}_{i} \left(\tilde{t}, f \right) = \frac{1}{\alpha + \Gamma G \mu} \left[ \frac{2k}{f} \frac{a(\tilde{t})}{a(t_0)} + \Gamma G \mu \tilde{t} \right]. 
\end{eqnarray}
In determining the relic GW spectra with the frequency as given in Eq. \eqref{GW-spectra},
we have used \texttt{micrOMEGAs} \cite{Alguero:2023zol}
for computing the scale factor $a(t)$ which depends on the relativistic {\it d.o.f} of the universe
as well as we have used the inbuilt integration routine in \texttt{micrOMEGAs} 
for computing the integration.
\begin{figure}[t] 
\centering
\includegraphics[width=0.75\textwidth]{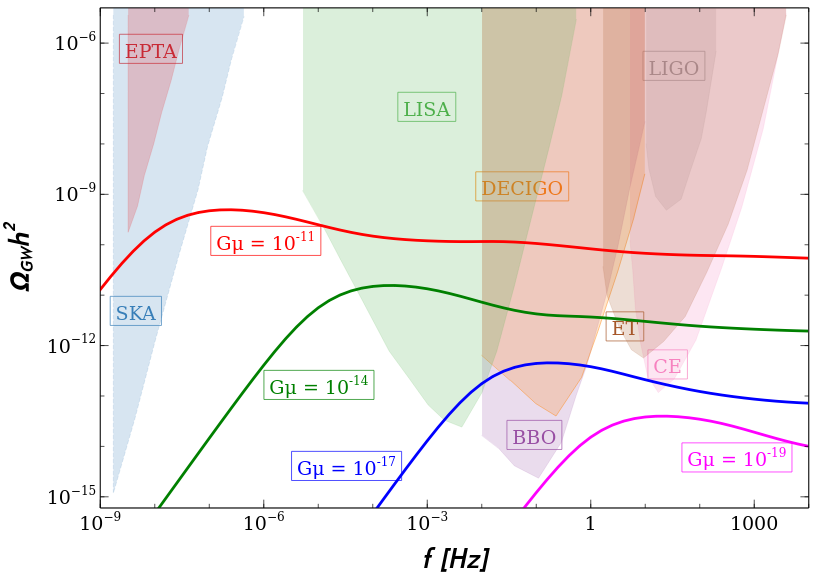} 
\caption{Variation of relic GW density with frequency for different values of string tension. Different colours represent the sensitivity prospects of various future GW detectors. EPTA data excludes cosmic string tensions $G\mu > 2 \times 10^{-11}$.} 
\label{fig:GW} 
\end{figure}

In Fig. \ref{fig:GW}, we show the prospects for GW produced 
from cosmic strings, as detectable by various proposed future detectors. 
In the plot, the low-frequency GW regime primarily originates from the matter-dominated era, while the high-frequency regime arises 
from the radiation-dominated era \cite{Cui:2018rwi}.
The string tension $G\mu > 2 \times 10^{-11}$ is already in tension with the 
EPTA data \cite{EPTA:2011kjn}, so we consider $G\mu < 10^{-11}$ in our analysis. This upper limit is already more stringent than 
the bound from the CMB, which is $G\mu < 2 \times 10^{-7}$ \cite{Charnock:2016nzm}.
As shown, different values of $G\mu$ will be probed by different proposed experiments. For example, $G\mu = 10^{-11}$ can be explored by all the proposed detectors, 
namely SKA \cite{Janssen:2014dka}, LISA \cite{LISA:2017pwj}, BBO/ DECIGO \cite{Yagi:2011wg}, 
ET \cite{Punturo:2010zz, Hild:2010id}, and CE \cite{LIGOScientific:2016wof}. 
On the other hand, $G\mu = 10^{-17}$ is beyond the reach of SKA but may be detectable by DECIGO and BBO, while values of $G\mu < 10^{-19}$ fall below the sensitivity limits of all future detectors.
The range of $G\mu$ accessible to GW detectors corresponds to a BSM 
scalar VEV in the range $v_{\phi} = 10^{10}$ GeV to $10^{14}$ GeV, 
which is also crucial for the KM3NeT signal, particularly in the scenario where dark matter decays 
are dominated by two-body standard model particles in the
final states, as discussed earlier.
Moreover, this range of $v_{\phi}$ can also yield the correct relic abundance of dark matter for $\mathcal{O}(100~\mathrm{PeV})$-scale DM masses through freeze-in mechanism, 
as will be discussed in more detail in Sec. \ref{sec:relic}. 
This provides an intriguing possibility of correlating freeze-in dark 
matter production with observable signals at GW detectors.

%----------------------------------------------------------------------------------
\section{Dark Mater Production}\label{sec:relic}
%----------------------------------------------------------------------------------
In this section, we study the relic density estimation
of the heavy DM using the freeze-in mechanism from ultraviolet 
dependent physics. For $\mathcal{O}(100)~ \mathrm{PeV}$ scale DM, 
the thermal freeze-out mechanism fails to generate the correct relic 
density due to the violation of unitarity bounds \cite{Griest:1989wd}. 
Therefore, we consider an alternative production mechanism, 
known as freeze-in \cite{Hall:2009bx}.
In our model, the right-handed (RH) neutrino is very heavy and beyond
the relevant scale of our study, so it can be integrated out safely. 
As a result, the interactions between the DM and other particles 
are mediated via dimension-5 operators. 
We consider a UV freeze-in mechanism for DM production where DM
relic density depends on the UV physics {\it e.g.} reheat temperature 
$T_R$. The squared amplitude for the DM production process 
$\phi \phi \rightarrow \bar \chi \bar{\chi}$ is given by: 
\begin{eqnarray} 
	\vert\mathcal{M}\vert^{2}_{\phi \phi \rightarrow \chi \bar{\chi} } = 2 \left(\frac{2\kappa^{2}}{M_{N}}\right)^{2} \left( s - 4 M^2_{\chi} \right) 
\end{eqnarray} 
and the corresponding cross section is: 
\begin{eqnarray} 
	\sigma_{\phi \phi \rightarrow \chi \bar{\chi}} = \frac{1}{8 \pi s} \left( \frac{2 \kappa^2}{M_{N}} \right)^{2} \left[ \frac{s - 4 M^2_{\chi}} {s - 4 M^2_{\phi}} \right]^{1/2} \left(s - 4 M^2_{\chi} \right) 
\end{eqnarray} 
where we assume $M_{N} \gg M_{\chi}, M_{\phi}$. 
DM production is dominated at very high temperatures, $T \gg M_{\chi}, M_{\phi}$, allowing us to safely neglect the masses of the particles compared to the center-of-mass energy $\sqrt{s}$. 
Under this approximation, the cross section simplifies to: 
\begin{eqnarray} 
	\sigma_{\phi \phi \rightarrow  \chi \bar{\chi} } = \frac{1}{8 \pi} \left( \frac{2 \kappa^{2}}{M_{N}} \right)^{2}. 
\end{eqnarray}
The Boltzmann equation governing DM production from pair or $\phi\phi$
annihilation is:
\begin{eqnarray} 
	\frac{d Y_{\chi}}{dz} = \frac{2 M_{sc} s(T)}{z^{2}  \mathcal{H}(T) T} \biggl[ \langle \sigma v \rangle_{\phi \phi \rightarrow \bar \chi \chi} Y^2_{\phi} - \langle \sigma v \rangle_{ \bar \chi \chi \rightarrow \phi \phi} Y^2_{\chi} \biggr], 
\end{eqnarray} 
where $z = M_{sc}/T$ and $M_{sc}$ is a free mass scale which 
can be taken as DM mass. 
The Hubble parameter is given by $ \mathcal{H}(T) = 1.66 \sqrt{g_{\rho}(T)} \frac{T^{2}}{M_{\rm pl}}$, and the entropy density 
is $s(T) = \frac{2 \pi^{2}}{45} g_{s}(T) T^{3}$, where $g_{\rho}(T)$
and $g_{s}(T)$ are the effective relativistic degrees of freedom 
of the Universe corresponding to the energy density and entropy 
density, respectively.
We assume that DM is produced via freeze-in from ultraviolet physics, with an initially negligible abundance. Hence, $Y_{\chi} = 0$ at early times, becoming nonzero once production starts from the thermal bath.
The thermally averaged cross section times velocity takes 
the form \cite{Gondolo:1990dk}, 
\begin{eqnarray} 
\langle \sigma v \rangle_{\phi \phi \rightarrow \chi \bar{\chi} } = \frac{g^2_{\phi} T}{2 (2\pi)^{4} n^2_{\phi} }\int~ds~ \sigma_{\phi \phi \rightarrow  \chi \bar{\chi} } (s-4 M^2_{\phi}) \sqrt{s} K_{1}\left( \frac{\sqrt{s}}{T} \right). 
\end{eqnarray} 
Since DM production occurs at very high temperatures, we can again neglect initial 
and final state masses. The thermal average then becomes:
\begin{eqnarray} 
\langle \sigma v \rangle_{\phi \phi \rightarrow \chi \bar{\chi}} \simeq \frac{g^2_{\phi} T}{2 (2\pi)^{4} n^2_{\phi}} \frac{1}{8 \pi} \left( \frac{2 \kappa^{2}}{M_{N}} \right)^{2} \int_{0}^{\infty} ds~ s^{3/2} K_{1}\left( \frac{\sqrt{s}}{T} \right). 
\end{eqnarray}
Using the above expression and $z = M_{sc}/T$, the Boltzmann equation 
can be rewritten as: 
\begin{eqnarray} 
	\frac{d Y_{\chi}}{d T} = - \frac{T^{5}}{ \mathcal{H}(T) s(T) (4 \pi^{5})} \left( \frac{2 \kappa^{2}}{M_{N}} \right)^2. 
\end{eqnarray}
Substituting expressions for $H(T)$ and $s(T)$, and integrating from the reheating temperature $T_{R}$ down to the present-day temperature $T_0$, the comoving number density is: 
\begin{eqnarray} Y_{\chi} &\simeq& \frac{3.6 M_{\rm pl} (T_{R} - T_{0})}{\pi^{7} g_{s} \sqrt{g_{\rho}}} \left( \frac{2 \kappa^{2}}{M_{N}} \right)^{2},\nonumber \\
	\label{Ychi} 
\end{eqnarray}
Once we calculate the co-moving number density, the resulting DM relic density 
is then given by \cite{Edsjo:1997bg},
\begin{eqnarray}
	\Omega_{\chi} h^{2} = 2.755 \times 10^{14} \left( \frac{M_{\chi}}{\PeV} \right) Y_{\chi}. 
	\label{omegaDM} 
\end{eqnarray}

We vary the model parameters over the following ranges: 
\begin{eqnarray} 
	10^{-7} \leq \kappa \leq &&1,\quad 10^{-20} \leq G\mu \leq 10^{-10},\quad 10^{10} \leq M_{N} [\mathrm{GeV}] \leq 10^{16},
	10^{-30} \leq y \leq 10^{-20},\nonumber 
	\\ && 1.7\times 10^{8} \leq M_{\chi}  [\mathrm{GeV}] \leq 5.5 \times 10^{9},\quad 10^{10} \leq T_{R} [\mathrm{GeV}] \leq 10^{13}. 
	\label{range-parameter}
\end{eqnarray}
We consider a DM mass range that can explain the KM3NeT events and apply the decay lifetime constraint $3.6 \times 10^{29} \leq \frac{\tau_{\chi}}{f_{\chi}} \leq 6.3 \times 10^{29}$, 
where $\tau_{\chi}$ is the DM lifetime and $f_{\chi}$ is its fractional contribution to the total DM abundance (we consider $1$–$100\%$). In scenarios where DM is a subdominant component,
a multi-component DM framework may be invoked, as explored in the literature \cite{Khan:2024biq}. 
As shown above, the right-handed neutrino mass $M_N$ 
is always greater than the reheating temperature $T_R$, meaning $N_R$ is not thermally populated 
in the early Universe. 
Usually, freeze-in DM cannot be probed by ongoing direct, indirect detection, or collider searches.
However, thanks to the KM3-230213A event, we might have a hint for decaying heavy DM produced via freeze-in process.
In our model, we can find solution for  both DM relic density and KM3-230213A event without
taking unnatural values for the relevant parameters in the model.
Also, gravitational wave is generated in the early Universe due to large Dark Higgs VEV. 
It has detection prospects at future gravitational wave detectors.

\begin{figure}[t]
\centering
\includegraphics[width=0.48\textwidth]{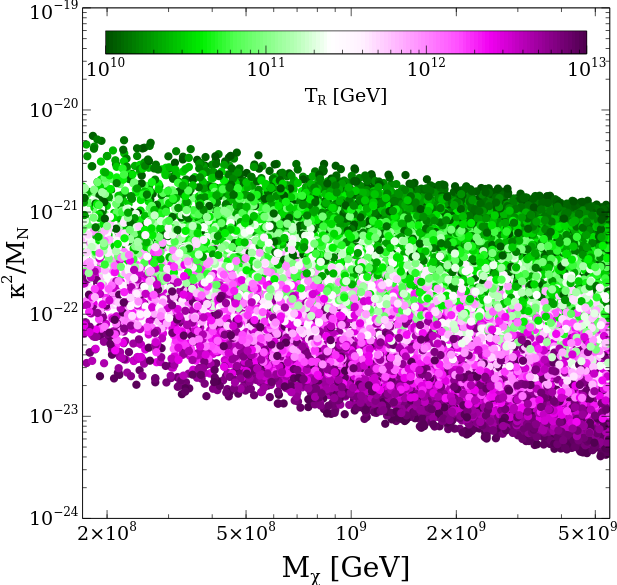}
\includegraphics[width=0.48\textwidth]{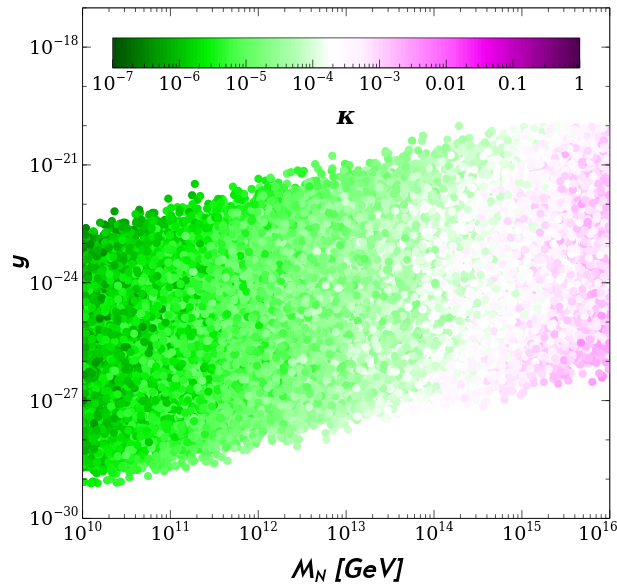} 
\caption{In the left panel (LP), we show the scatter plot in the $(M_{\chi},~\frac{\kappa^{2}}{M_{N}})$ plane, whereas the right panel (RP) displays the scatter plot in the $(M_{N},~y)$ plane.
The color gradient in the LP represents different values of the reheating temperature $T_R$,
while in the RP, it corresponds to different values of the coupling $\kappa$.
The other parameters, which are not shown, have been varied as listed 
in Eq.~\eqref{range-parameter}. 
}
\label{fig:dm1}
\end{figure}

In the LP of Fig.~\ref{fig:dm1}, we show the scatter plot in the 
$(M_{\chi},~\frac{\kappa^{2}}{M_{N}})$ plane after imposing the 
requirement that the DM mass and its fractional contribution 
lie within the $(1$–$100)\%$ range, providing the correct value 
of $\frac{\tau_{\chi}}{f_{\chi}}$ necessary to explain the 
KM3NeT signal.
An anti-correlation between $M_{\chi}$ and 
$\frac{\kappa^{2}}{M_{N}}$ is observed, which can be understood 
from Eqs.~(\ref{Ychi}) and (\ref{omegaDM}). 
These equations show that the DM relic density is proportional 
to $M_{\chi} T_{R} \left( \frac{2 \kappa^{2}}{M_{N}} \right)^2$. 
Therefore, to achieve a fixed relic density, an increase in one 
parameter must be compensated by a decrease in the other, 
leading to the observed anti-correlation.
From the color bar, we also observe that as $\frac{\kappa^{2}}{M_N}$ increases, lower values of $T_R$ are required (as seen along the y-axis). Similarly, along the x-axis, as $M_{\chi}$ increases, the required $T_R$ also decreases to maintain the correct relic density.
In the RP, we present the scatter plot in the $(M_{N}~,y)$ plane, with the color bar representing different values of $\kappa$. 
The parameters in this plot are related to both the DM relic 
density and decay width. As $M_{N}$ increases, the relic density 
tends to decrease, while it increases with larger $\kappa$. 
On the other hand, the decay width increases with increasing $y$. 
To obtain the correct value of $\frac{1}{f_{\chi} \Gamma_{\chi}}$ for explaining the KM3NeT signal, we observe a correlated variation between $M_{N}$ and $y$: an increase in $M_{N}$ lowers $f_{\chi}$, while an increase in $y$ raises $\Gamma_{\chi}$, keeping the product $f_{\chi} \Gamma_{\chi}$ approximately constant.
Furthermore, to obtain the correct DM relic density, $M_{N}$ and $\kappa$ must also vary in a correlated (roughly linear) manner, as indicated by the color gradient.

\begin{figure}[t]
\centering
\includegraphics[width=0.48\textwidth]{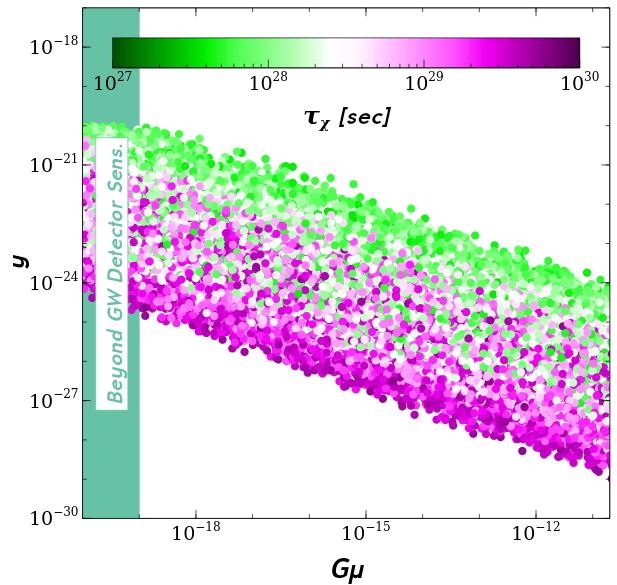}
\includegraphics[width=0.48\textwidth]{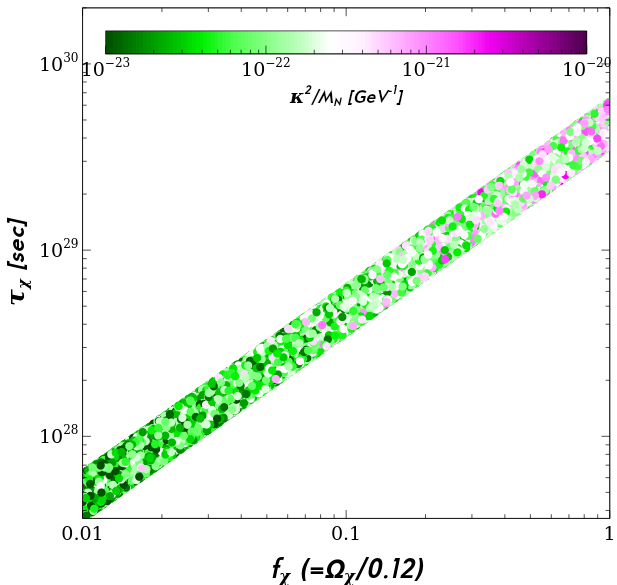} 
\caption{The LP and RP show scatter plots in the 
$(G\mu,~y)$ and $(f_{\chi},~\tau_{\chi})$ planes, respectively. 
In the LP, the color gradient represents different values 
of $\tau_{\chi}$, while in the RP, it corresponds to values 
of $\kappa^{2}/M_{N}$. }
\label{fig:dm2}
\end{figure}
In the LP and RP of Fig. \ref{fig:dm2}, we show the allowed regions in the $(G\mu,~y)$ and $(f_{\chi},~\tau_{\chi})$ planes, respectively. The color variation in the LP indicates different values of the DM lifetime $\tau_{\chi}$, while in the RP, it corresponds to the values of $\kappa^{2}/M_{N}$. In the LP, we observe an anti-correlation between $G\mu = (v_{\phi}/M_{\rm pl})^{2}$ and the Yukawa coupling $y$. This is because the decay width is proportional to $(y v_{\phi})^{2}$; thus, for larger values of $G\mu$, a smaller value of $y$ is required to achieve the correct decay width that explains the KM3NeT signal. Consequently, since the DM lifetime $\tau_{\chi} = 1/\Gamma_{\chi} \propto 1/(y v_{\phi})^{2}$, we find that $\tau_{\chi}$ decreases with increasing $G\mu$ and also with increasing $y$, consistent with the color gradient in the plot. The cyan region in the LP corresponds to $G\mu < 10^{-19}$, which lies below the sensitivity of future GW detectors such as BBO and hence cannot be probed. However, the other parameter values fall within the reach of proposed GW detectors, as discussed in Sec. \ref{sec:GW}.
In the RP, we display the scatter plot in the $(f_{\chi},~\tau_{\chi})$ plane, where a sharp correlation emerges, reflecting the requirement to explain the KM3NeT signal. The DM relic density scales as $\Omega_{\chi} h^{2} \propto (\kappa^{2}/M_{N})^{2}$, while the DM lifetime follows $\tau_{\chi} \propto 1/(\kappa/M_{N}^{2})$. Since both relic density and lifetime also depend on other parameters such as the DM mass $M_{\chi}$, Dirac Yukawa coupling $y$, and the VEV $v_{\phi}$, we observe a mixed color distribution—e.g., combinations of green and magenta points—within the same region of parameter space. As mentioned earlier, all the points shown in Figs. \ref{fig:dm1} and \ref{fig:dm2} are consistent with the KM3NeT signal, satisfy the allowed relic density range, and lie within the reach of future gravitational wave detectors.

%----------------------------------------------------------------------------------
\section{Conclusion}\label{sec:conclusion}
%----------------------------------------------------------------------------------
In this work, we explore dark $U(1)_X$ gauge symmetry.
This DM model can resolve the KM3-230213A event in terms of DM decay.
Thanks to the heavy RH neutrino portal interaction, we can naturally obtain very tiny couplings 
between DM and SM particles,  
inducing DM decays into SM particles, including active neutrinos, 
with the long enough lifetime.
In the case of $v_\phi \gg m_\chi$, the dominant DM decay channels are 
$\chi \to \nu h, \nu Z, \ell^\pm W^\mp$ but for the opposite regime $v_{\phi} \ll m_{\chi}$ 
DM dominantly decays to three-body final states, which case has not been explored in the present study. 
To explain the KM3NeT signal, we have considered DM mass at $m_{\chi} = 440$ PeV and 
its lifetime $\tau_{\chi} = 5 \times 10^{29}$ sec as the benchmark point.
It should be noted that the present study can alleviate the tension between the non-observation at IceCube and the observation in KM3NeT data, which could exceed $3\sigma$ depending on the sources considered, by suitably choosing the DM mass and lifetime.
In calculating the neutrino flux, we have considered both galactic and 
extragalactic contributions in determining the neutrino flux and 
found the peak at the neutrino energy 
$E_{\nu} = 220$ PeV which can explain the KM3-230213A data. In determining the 
extragalactic contribution we have taken into account the redshift effect due the 
matter dilution, dimming of sources and proper length. Moreover, we have also
considered neutrino opacity in determining the extragalactic contribution which 
reduced the flux strength by $20\%$. Additionally, we have also estimated the 
photon flux predicted by our work which are below the LHAASO-KM2A and EAS-MSU data.
In determining the neutrino flux as well 
as the photon flux, we have used the HDMspectra. 

As said we have focussed on $v_{\phi} \gg m_{\chi}$ so that the DM decays dominantly into two SM particles,
which results in the cosmic string production in the early in the detectable range.
We have found that cosmic strings tension in the range $G\mu = 10^{-11}$
to $10^{-19}$ (corresponds to $v_{\phi} \sim 10^{14}$ GeV to $10^{8}$ GeV)
can be detected at the different proposed experiments like SKA, BBO/DECIGO,  
LISA, ET and CE. The same set of $G\mu$ range can predict the DM 
mass and lifetime in the correct range which can explain the KM3NeT data.
In estimating the GW spectra in particular the scale factor and 
numerical integration we have used the popular package \texttt{micrOMEGAs}.
Finally, we have discussed in detail the PeV-scale DM production using the UV
freeze-in mechanism. We have $\phi \phi \rightarrow \bar \chi \chi$
process, suppressed by the heavy right-handed neutrino mass, which is dimension-5
operator and produces DM in the early Universe dominantly. 
We have found that $T_{R} \sim 10^{10}$ GeV, $\kappa \sim 10^{-4}$, $M_{N} \sim 10^{12}$,
$v_{\phi} = 10^{13}$ GeV and $y \sim 10^{-20}$, 
we can produce the DM in the correct range and the low value of $y$ is needed 
to make the DM decay lifetime $\mathcal{O}(10^{29})$ sec.
We have shown a few scatter plots which can predict DM in the $(1-100)\%$
range and at the same time can explain the KM3NeT signal and future possibility
to detect at different GW detectors.

\acknowledgments
The work is supported in part by Basic Science Research Program through the National Research Foundation of Korea (NRF) funded by the Ministry of Education, Science and Technology (NRF-2022R1A2C2003567 (SK, JK) and RS-2024-00341419 (JK)) and by KIAS Individual Grants under Grant No. PG021403 at Korea Institute for Advanced Study (PK).
The work is also supported by Brain Pool program funded by the Ministry of Science and ICT through the National Research Foundation of Korea (RS-2024-00407977) (SK).
This work used the Scientific Compute Cluster at GWDG, the joint
data center of Max Planck Society for the Advancement of Science (MPG) and University of
G\"{o}ttingen.

\appendix
%%%%%%%%%%%%%%%%%%%%
\section{Dark Matter Decay Width}
\label{dm-decay}
%%%%%%%%%%%%%%%%%%%%
The DM $\chi$ can decay to $\nu h$, $\nu Z$, and $l^{\pm}W^{\mp}$ through the vertices shown in Eq.~\eqref{eq:operator}. The decay widths can be expressed in the limit $M_{\chi} \gg M_{Z,W,l,h}$ as follows,
\begin{eqnarray}
\Gamma_{\chi \rightarrow \nu h} \simeq \frac{M_{\chi}}{32 \pi} \left( \frac{y \kappa v_{\phi}}{2 M_{N}} \right)^{2},~~
\Gamma_{\chi \rightarrow \nu Z} \simeq \theta^{2} \frac{M^3_{\chi}}{32\pi v^2_{H}},~~
\Gamma_{\chi \rightarrow l^{\pm} W^{\mp}} \simeq \theta^{2} \frac{M^3_{\chi}}{16\pi v^2_{H}},
\end{eqnarray}
where the mixing angle $\theta = \frac{y \kappa v_{\phi} v_{H} }{2 M_{N} M_{\chi}}$, as defined in Eq.~\eqref{mixing-angle}. It is to be noted that the decay widths follow the ratio $\Gamma_{\chi \rightarrow \nu h} : \Gamma_{\chi \rightarrow \nu Z} : \Gamma_{\chi \rightarrow l^{\pm}W^{\mp}} = 1:1:2$, which is also predicted by the Goldstone equivalence theorem. 

\bibliographystyle{JCAP}
\bibliography{ref}
\end{document}